\documentclass[letterpaper,twocolumn,showpacs,prl,superscriptaddress,floatfix]{revtex4}
\usepackage[pdftex]{graphicx}
\usepackage{amssymb}
\usepackage{amsmath}
\usepackage{textcomp}
\usepackage{times}
\begin{document}

\date{\today} 

\title{Anisotropic softening of magnetic excitations in lightly electron doped Sr$_2$IrO$_4$}
\author{X. Liu}
\email{xliu@iphy.ac.cn} 
\affiliation{Beijing National Laboratory for Condensed Matter Physics and Institute of Physics, Chinese Academy of Sciences, Beijing 100190, China}
\affiliation{Collaborative Innovation Center of Quantum Matter, Beijing, China}
\affiliation{Condensed Matter Physics and Materials Science Department, Brookhaven National Laboratory, Upton, New York 11973, USA}

\author{M. P. M. Dean}
\affiliation{Condensed Matter Physics and Materials Science Department, Brookhaven National Laboratory, Upton, New York 11973, USA}

\author{Z. Y. Meng} 
\affiliation{Beijing National Laboratory for Condensed Matter Physics and Institute of Physics, Chinese Academy of Sciences, Beijing 100190, China}
 
\author{M. H. Upton}
\affiliation{Advanced Photon Source, Argonne National Laboratory, Argonne, Illinois 60439, USA}

\author{T. Qi}
\affiliation{Center for Advanced Materials, Department of Physics and Astronomy, University of Kentucky, Lexington, Kentucky 40506, USA}

\author{T. Gog}
\affiliation{Advanced Photon Source, Argonne National Laboratory, Argonne, Illinois 60439, USA}

\author{Y. Cao}
\affiliation{Condensed Matter Physics and Materials Science Department, Brookhaven National Laboratory, Upton, New York 11973, USA}

\author{J. Q. Lin}
\affiliation{Beijing National Laboratory for Condensed Matter Physics and Institute of Physics, Chinese Academy of Sciences, Beijing 100190, China}

\author{D. Meyers}
\affiliation{Condensed Matter Physics and Materials Science Department, Brookhaven National Laboratory, Upton, New York 11973, USA}

\author{H. Ding}
\affiliation{Beijing National Laboratory for Condensed Matter Physics and Institute of Physics, Chinese Academy of Sciences, Beijing 100190, China}
\affiliation{Collaborative Innovation Center of Quantum Matter, Beijing, China}

\author{G. Cao}
\affiliation{Center for Advanced Materials, Department of Physics and Astronomy, University of Kentucky, Lexington, Kentucky 40506, USA}

\author{J. P. Hill}
\affiliation{Condensed Matter Physics and Materials Science Department, Brookhaven National Laboratory, Upton, New York 11973, USA}

\def\mathbi#1{\ensuremath{\textbf{\em #1}}}
\def\Q{\ensuremath{\mathbi{Q}}}

\begin{abstract}
{The magnetic excitations in electron doped (Sr$_{1-x}$La$_x$)$_2$IrO$_4$ with $x = 0.03$ were measured using resonant inelastic X-ray scattering at the Ir $L_3$-edge. Although much broadened, well defined dispersive magnetic excitations were observed. Comparing with the magnetic dispersion from the undoped compound, the evolution of the magnetic excitations upon doping is highly anisotropic. Along the anti-nodal direction, the dispersion is almost intact. On the other hand, the magnetic excitations along the nodal direction show significant softening. These results establish the presence of strong magnetic correlations in electron doped (Sr$_{1-x}$La$_x$)$_2$IrO$_4$ with close analogies to the hole doped cuprates, further motivating the search for high temperature superconductivity in this system.}
\end{abstract}


\maketitle

Together with the tremendous research activity on the superconducting cuprates \cite{PALee2006, BKeimer2015}, efforts to compare the cuprates with other related systems have also been on-going for decades. Such comparison serves as a natural approach to clarify the roles of multiple emergent phenomena in the phase diagram of the cuprates, including magnetic fluctuations, superconductivity, pseudo gap and charge density waves {\it etc}. The $5d$ oxide Sr$_2$IrO$_4$ is an excellent candidate for such study. This so called spin-orbit-coupling driven Mott insulator \cite{BJKim2008} is in close proximity to the single layered cuprate La$_2$CuO$_4$, both structure-wise  \cite{MKCrawford1994} and electronically \cite{YKKim2014,YKKim2015,ATorre2015,YCao2014}. Sr$_2$IrO$_4$ hosts a single hole in the $t_{2g}$ manifold where a Mott gap is opened, assisted by strong spin-orbit coupling \cite{BJKim2008, BJKim2009, HJin2009}, and its magnetic excitation spectrum can be well described using a Heisenberg model of effective spin-$\frac{1}{2}$ moments on a square lattice \cite{JHKim2012,LJPAment2011}.  With a minimum single band model, Sr$_2$IrO$_4$ and La$_2$CuO$_4$ are strikingly similar \cite{FWang2011}, leading to the proposal that this compound could also host unconventional high temperature superconductivity (HTS) upon doping \cite{FWang2011, ZYMeng2014}. Moreover, due to the opposite signs of the next-nearest-neighbor hopping integral in these two systems, theoretical work further suggests that the electron doped Sr$_2$IrO$_4$ might be more closely analogous to those of hole (rather than electron) doped cuprates \cite{FWang2011, ZYMeng2014}.

Although the phase diagram of doped Sr$_2$IrO$_4$ has not been fully explored, a large amount of experimental work supports the hypothesis that the Fermiology of the doped iridates is closely analogous to the cuprates. Upon doping with La up to $6\%$, Sr$_2$IrO$_4$ evolves from an antiferromagnetically ordered insulator to a paramagnetic \cite{MGe2011} or percolative \cite{XChen2015} metal. A T-linear resistivity was observed with potassium substitution \cite{MGe2011}. Further, angle-resolved photoemission spectroscopy (ARPES) data from several groups \cite{YKKim2014, YKKim2015, ATorre2015, YCao2015} has shown convincingly that doping indeed drives a similar low energy electron evolution as observed in the cuprates.  An anisotropic pseudo gap opens on the Fermi surface, with the same symmetry as that of the cuprates. However, the related question of whether the magnetic correlations that are often implicated in HTS \cite{DJScalapino2012, FWang2011, ZYMeng2014} are also analogous remains largely unexplored. 

Here we measure the magnetic excitations in electron doped (Sr$_{1-x}$La$_x$)$_2$IrO$_4$ with $x=0.03$ using Ir $L_3$ edge resonant inelastic X-ray scattering (RIXS). The samples at this doping show weakly metallic behavior and a strongly reduced magnetic moment induced by field \cite{MGe2011,undoped}. In the ARPES measurements, a sizable Fermi surface \cite{ATorre2015, YCao2015} was observed. The phase diagram summarized by X.~Chen {\it et al.} \cite{XChen2015} suggests that this doping sits at the border of the metal-insulator transition. We observe that doping induces damped magnetic excitations with a strongly anisotropic softening compared to the undoped compound. Along the anti-nodal direction, the dispersion is almost intact with doping. Along the nodal direction, however, the magnetic excitations are strongly softened by about $~27\%$. This phenomenology is closely anaologous to the hole doped cuprates and supports theoretical notions that electron-doping iridates may induce HTS \cite{FWang2011, ZYMeng2014}.

The high quality single crystals used were synthesized using a self-flux technique \cite{MGe2011}. Ir $L_3$-edge RIXS measurements were carried out at 27ID and 30ID of the Advanced Photon Source. The incident X-ray energy was set to 11.215~keV based on a resonant condition survey. The overall experimental resolution was about 37~meV (FWHM) estimated from a Lorentzian fitting of the quasi-elastic line. All data presented were collected at low temperature of 20~K. For convenience, the reciprocal space here is indexed using $I4/mmm$ notation, where the $[1,0,0]$ direction is parallel to the nearest Ir-Ir bond directions and, in analogy to the notation used in electronic structure measurements \cite{YKKim2014, YKKim2015, ATorre2015,ADamascelli2003} we refer to this as the anti-nodal direction, whereas the Brillouin zone diagonal $[1,1,0]$ is referred to as the nodal direction. The insensitivity of the RIXS spectra to the out-of-plane direction was confirmed at multiple \Q{} points with different out-of-plane momentum transfer values, consistent with the 2D nature of this material. 

\begin{figure}[h]
\includegraphics[width=0.45\textwidth]{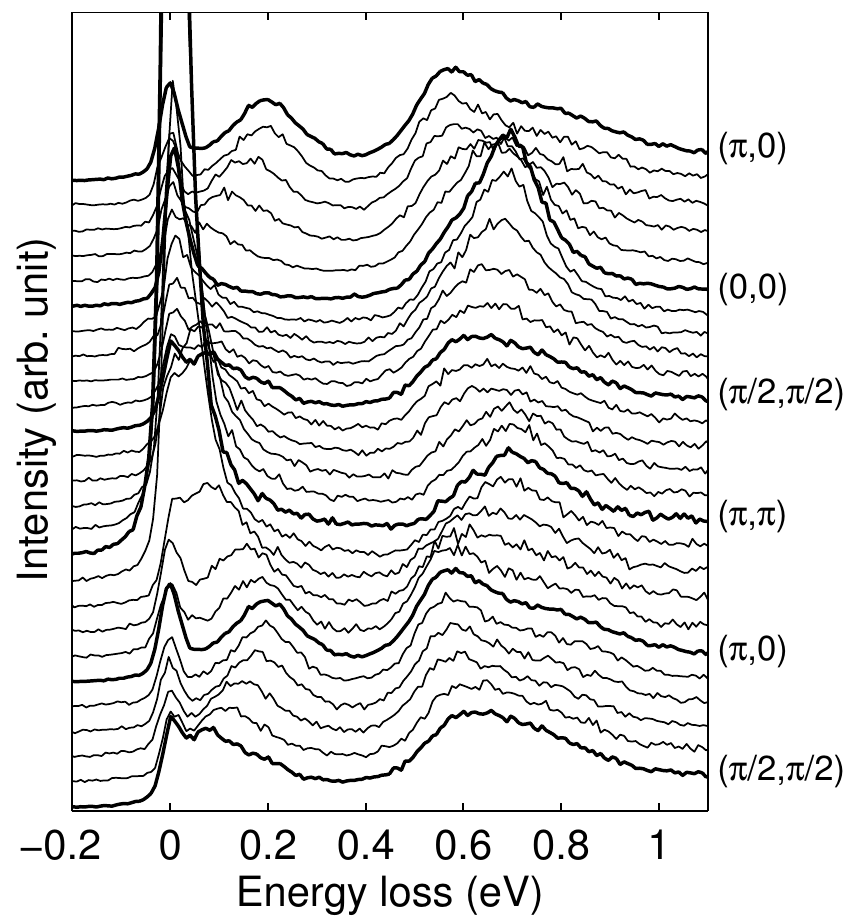}
\caption{RIXS energy loss spectra recorded along several cuts in reciprocal space. A damped magnon excitation is observed dispersing from low energy at $(0,0)$ and $(\pi,\pi)$ to higher energies around the zone boundary at $(\pi,0)$ and $(\pi/2,\pi/2)$. The spin-orbit exciton mode is seen around 0.7~eV.}
\label{stack}
\end{figure}

Figure \ref{stack} summarizes the RIXS spectra taken at different \Q{} points along several cuts in reciprocal space. In addition to the quasi-elastic lines centered at zero energy loss, two dispersive features can be observed below 1~eV. Such an observation is similar to that on the undoped compound \cite{JHKim2012, JHKim2014}. The higher energy dispersive feature (centered around 0.7~eV) was suggested to originate from the so called spin-orbit exciton, and the lower energy dispersion was assigned as magnetic in nature. The magnetic dispersion maximum reaches approximately 200~meV at the $(\pi, 0)$ point, similar to the magnon bandwidth obtained on the undoped sample. The relative strength of the magnetic excitations to the spin-orbit exciton excitations observed here is similar to that observed in the undoped compound with the same experimental geometry. Since the creation of the exciton is expected to scale with the unoccupied $t_{2g}$ states, the doping could result in no more than 6\% reduction of its spectral weight. Based on this argument, we conclude that the spectral weight of the magnetic excitations are not significantly suppressed upon doping.  

The dispersion of the observed magnetic excitations was extracted by fitting the RIXS spectra up to 0.45~eV energy loss. The quasi-elastic line was fitted with a Lorentzian, whose width was pre-determined as 37~meV FWHM from an off-resonant spectrum and kept fixed during the fitting. The upturn on the high energy loss side was accounted for with a Gaussian tail. For the magnetic excitation, we model the imaginary part of the magnetic susceptibility with a Lorentzian and note that the measured RIXS intensity, $I(\Q{}, \omega)$, is proportional to this multiplied by the Bose-Einstein factor, 
\begin{equation}
I(\Q{}, \omega) \propto \frac{\Gamma_{\Q}}{(\omega - \omega_{\Q})^2 + \Gamma^2_{\Q{}}} \cdot \frac{1}{1 - e^{-\omega / k_B T}} ,
\end{equation} 
where $\omega_{\Q}$ and $\Gamma_{\Q{}}$ are the energy and FWHM of the magnetic excitation at \Q{} respectively, and $k_B$ is the Boltzmann constant. This form was numerically convolved with a Lorentzian in order to account for the experimental resolution. Similar formalism was successfully applied to the doped cuprates \cite{MTacon2011, MDean2013, MDean2015}. 

\begin{figure}[h]
\includegraphics[width=0.48\textwidth]{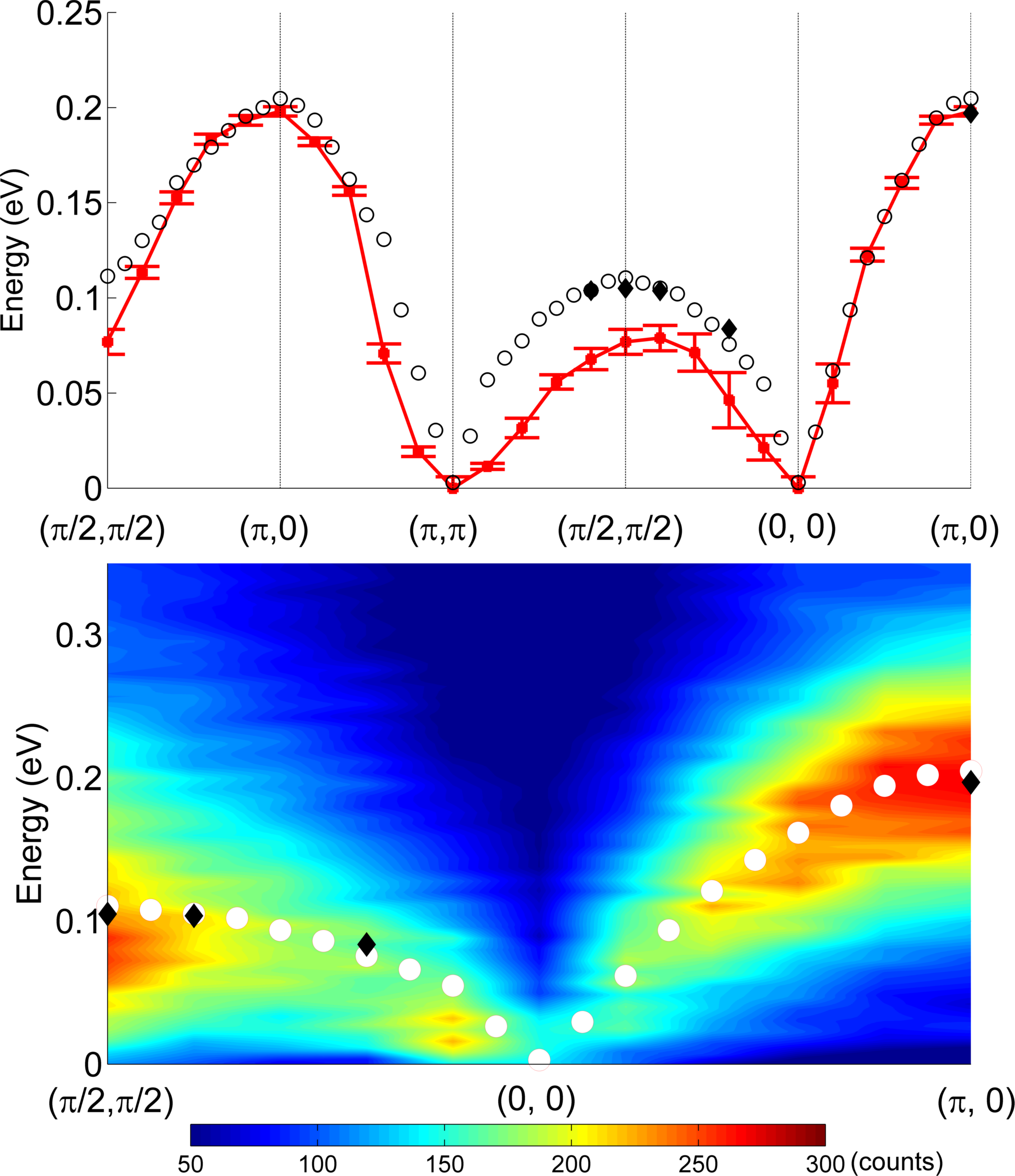}
\caption{A comparison of the dispersion of the magnetic excitations in (Sr$_{1-x}$La$_x$)$_2$IrO$_4$ with $x = 0.03$ with the undoped compound. Top pannel: the solid squares  and diamonds are from the current work on the doped and undoped samples, and the open circles are from the undoped compound extracted from Ref.~\cite{JHKim2012}. Bottom pannel: the energy-momentum intensity map of the measured RIXS spectra with the quasi-elastic line subtracted; the white dots(Ref.\cite{JHKim2012}) and black diamonds (our sample) are for the undoped compound.}
\label{dispersion}
\end{figure}

The obtained dispersion relation is shown in the top panel of Fig~.\ref{dispersion} (solid red squares), together with that from the undoped compound. The open circles are the data extracted from Ref.~\cite{JHKim2012}, and the black diamonds are the magnon energies measured on our own undoped sample \cite{undoped}. Upon $3\%$ doping, the magnetic excitations in (Sr$_{1-x}$La$_x$)$_2$IrO$_4$ are modified in a strongly anisotropic manner. Along the $(0, 0)\rightarrow(\pi, 0)$ anti-nodal direction, the magnetic dispersion in the doped sample follows that of the undoped compound closely. The maximum at $(\pi, 0)$ zone boundary determines a dispersion bandwidth of $\sim200$~meV. In contrast, there is a significant ``softening'' along the $(0, 0)\rightarrow(\pi, \pi)$ nodal direction with doping. At the $(\pi/2, \pi/2)$ zone boundary point, the magnetic excitation is softened by about 27\%. Such anisotropic response can be seen more directly in the energy-momentum intensity map in the bottom panel of Fig~.\ref{dispersion} where the measured RIXS spectra with only the quasi-elastic line subtracted are shown. The comparison presented in Fig.~\ref{dispersion} leads to our two major observations. Firstly, the magnetic excitations from the isospin-$\frac{1}{2}$ square lattice in the undoped compound evolve, but clearly survive upon $3\%$ doping where a sizable Fermi-surface has well developed \cite{ATorre2015, YCao2015}. Secondly, the response of the magnetic excitations to doping is highly anisotropic.

\begin{figure}[h]
\includegraphics[width=0.45\textwidth]{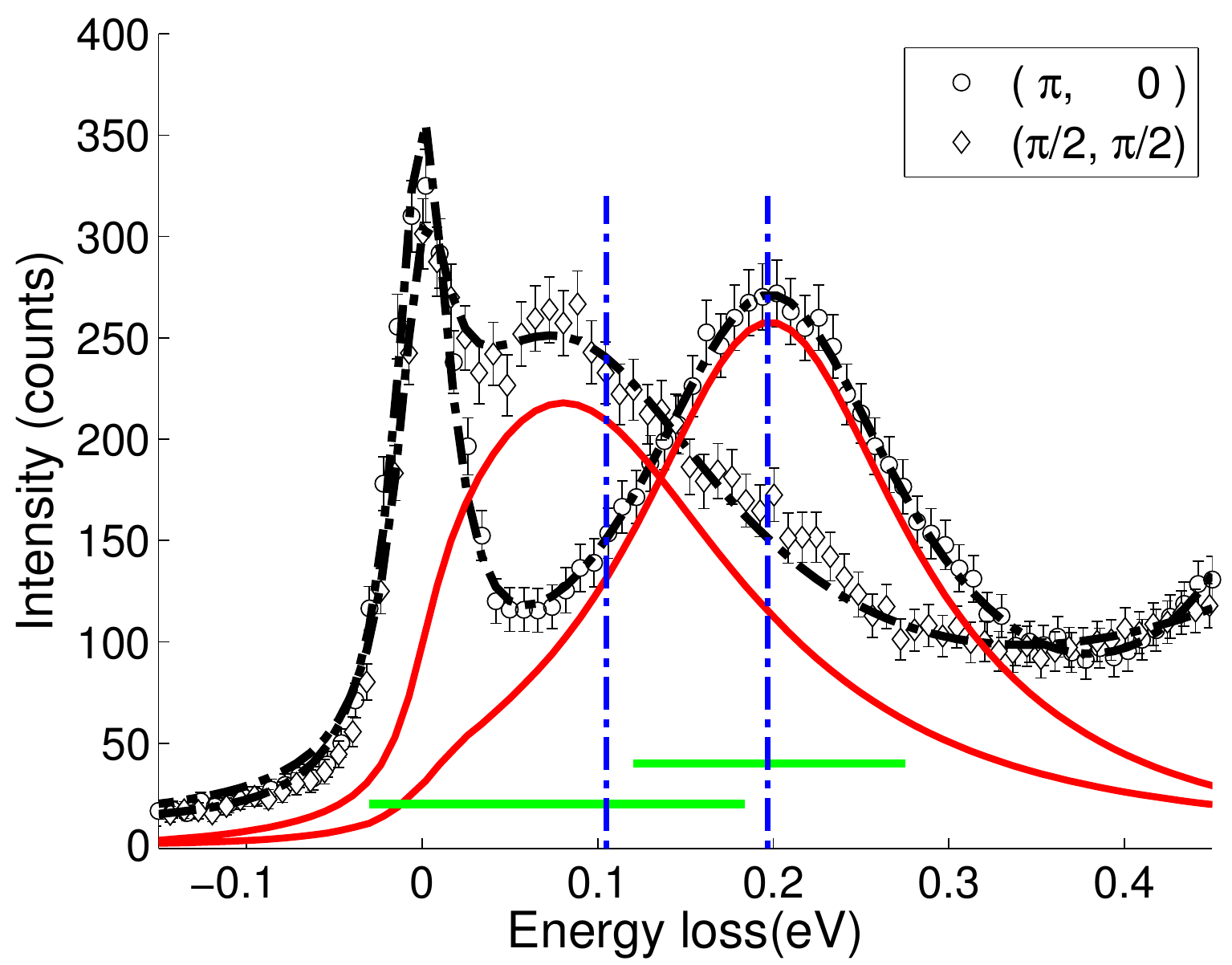}
\caption{RIXS spectra for $(\pi/2, \pi/2)$ (open diamonds) and $(\pi, 0)$ (open circles) respectively. Dashed lines through the data points show our fitting results. The two peaks depicted by solid red lines are the fitted magnetic excitation peaks, with horizontal bars showing their FWHM of the Lorentzian model in Eq.~1. The vertical dashed lines indicate the magnon energies at these two \Q{} points in our undoped sample \cite{undoped}.}
\label{comp2Q}
\end{figure}

More details of such anisotropic impact from doping can be seen in Fig.~\ref{comp2Q} where the RIXS spectra for two characteristic zone boundary points, namely $(\pi/2, \pi/2)$ and $(\pi, 0)$, are plotted. The data was fitted with the scheme described earlier. The faster down turn across zero towards the negative energy loss side of the fitted magnetic excitation peaks (red solid lines) is due to the Bose-Einstein factor. For comparison, the magnon energies of our undoped sample are indicated with vertical dashed lines.  

At both \Q{} points, the magnetic excitations are much broader than the instrumental resolution. This is in contrast to the case of the undoped compound \cite{JHKim2012,LJPAment2011,undoped,JHKim2014}, for which the magnons are resolution limited. Thus, the magnetic excitation is highly damped upon doping. From our fitting, the center of the magnetic excitation at $(\pi, 0)$ agrees well with that of the undoped compound. For $(\pi/2, \pi/2)$, the large softening is clearly observable from the raw experimental data. Our fitting gave a peak center of $77(\pm6)$~meV, compared to 110~meV from Ref. \cite{JHKim2012} and 105~meV from our own undoped sample. Such anisotropic impact from doping also appears in the broadening of the magnetic excitations. The fitted widths (FWHM) are 155 and 214~meV for $(\pi, 0)$ and $(\pi/2, \pi/2)$ respectively. The more broadened magnetic excitations suggest stronger scattering along the nodal direction.

The anisotropic response to doping implies that our observed magnetic excitations are {\it not} from macroscopic phase separation, which might be a general concern in doped systems. Rather, the introduced carriers must either co-exist with or reside in nano-scale proximity to the observed magnetism. On the other hand, we do not rule out microscopic inhomogeneity, which has been commonly observed in electron correlated systems upon doping \cite{McElroy2005}. Indeed, with scanning tunneling microscopy measurements on their $5\%$ doped sample, X.~Chen {\it et al.} \cite{XChen2015} observed nano-scale inhomogeneous distribution of the local density of states, which is further supported by the work by Y.~J.~Yan {\it et al.} \cite{YJYan2015}. Similarly, nano-scale inhomogeneity has also been observed in hole-doped cuprates \cite{McElroy2005}, and might be an intrinsic feature of these doped Mott insulators. 

The doping evolution of the magnetic excitations we report here shares much in common with that has been observed in the hole doped cuprates. It has been shown \cite{MTacon2011, MDean2013, MDean2015, MDean2013b} that the magnetic excitations survive into the heavily overdoped region in many cuprate families with both dispersions and spectral weights similar to those of undoped compounds. For Sr$_2$IrO$_4$, the ARPES measurements \cite{ATorre2015, YCao2015} show that doping significantly reforms the $J_{\text{eff}}=\frac{1}{2}$ band in a few hundreds meV range. At the doping level studied here, the $J_{\text{eff}}=\frac{1}{2}$ band has already extended across the Fermi energy, resulting in a sizable Fermi surface \cite{YCao2015}. The indirect charge gap seems to be closed with the appearance of the electron-like pockets around $(\pi/2, \pi/2)$ and hole-like pockets around $(\pi, 0)$. To explain such a drastic response, A.~de~la~Torre {\it et al.} \cite{ATorre2015} suggests that doping strongly weakens the onsite Coulomb repulsion, leaving (Sr$_{1-x}$La$_x$)$_2$IrO$_4$ to be in the weakly interacting region. Such strong doping dependence of the Coulomb U was also discussed for cuprates \cite{CKusko2002}, which was challenged by many others \cite{OnSiteU}. Our results show that, although the doping significantly renormalizes the electron band structure in (Sr$_{1-x}$La$_x$)$_2$IrO$_4$, the system still supports magnetic correlations with similar dispersion and spectral intensity, indicating that it has a similar degree of electron-electron correlation as the undoped compound.    

The observation of anisotropic modification to the magnetic excitations is particularly interesting. We notice that such anisotropic softening of magnetic excitations along the nodal direction has been also observed with RIXS in hole-doped superconducting cuprates as well \cite{MGuarise2014, MDean2014}, where in doped Bi-based cuprates, the magnons collapse along the nodal direction but persist along the anti-nodal direction in the momentum space. Such behavior is captured in random phase approximation (RPA) calculations of the magnetic response in the cuprates based on itinerant quasiparticles \cite{MGuarise2014, MDean2014, AJames2012, RZeyher2013, MEremin2013}. Hence, such anisotropic softening of the magnetic excitations in doped cuprates was associated with the emergent correlated itinerant, or partially itinerant, nature of the electrons in the nodal direction close to $(\pi/2, \pi/2)$ (i.e., the emergent Fermi arc). The spin fluctuations in the nodal direction strongly decay into the emergent electron-hole continuum and are hence softened and damped. 

Interestingly, such a picture is somehow more appropriate in the context of doped Sr$_2$IrO$_4$ than that in the doped cuprates. The Coulomb repulsion in the Ir $5d$ shell is weaker than that in the Cu $3d$ shell, and ARPES experiments \cite{ATorre2015, YCao2015} have reported that the Sr$_2$IrO$_4$ is much more sensitive towards doping than the cuprates. In fact, a low doping already generates a Fermi surface with coherent nodal excitations and an anti-nodal pseudogap. Hence, if it is the coupling to the correlated itinerant quasiparticles (electron-hole continuum associated with the emergent Fermi surface) that give rise to the anisotropic softening of magnetic excitations along the nodal direction in underdoped curprates, this coupling should be stronger and more likely to happen in doped Sr$_2$IrO$_4$, as the itinerant nature of the lightly doped Ir 5d electrons are more pronounced, and indeed, this is what has been clearly observed in the RIXS data presented in this work.

In conclusion, we have shown that 3\% doped (Sr$_{1-x}$La$_x$)$_2$IrO$_4$ still hosts persistent magnetic excitations with similar spectral intensity to the undoped compound, despite the fact that a sizable Fermi Surface has formed at this doping level \cite{ATorre2015, YCao2015}. At the same time, the observed magnetic excitations respond to doping anisotropically. Along the anti-nodal direction, the dispersion is almost intact with doping. On the other hand, along the nodal next nearest Ir-Ir direction, the magnetic excitations are strongly softened and more damped. These observations unambiguously point to a strong coupling between the doped electrons and the magnetism in this  system. Such a behavior is closely analogous to hole-doped cuprates, further motivating the search for high temperature superconductivity in this system.

\emph{Note added:} During our submission, we noticed a similar observation reported by H. Gretarsson and collaborators \cite{Gretarsson2016}.

We thank T. Xiang for fruitful discussions. X. L. acknowledges financial support from MOST (No. 2015CB921302) and CAS (Grant No: XDB07020200) of China. Z. Y. M. is supported by the National Natural Science Foundation of China (Grant Nos. 11421092 and 11574359). Both X. L. and Z. Y. M. are supported by the National Thousand Young-Talents Program of China. The work at Brookhaven was supported by the U. S. Department of Energy, Division of Materials Science, under Contract No. DE-AC02-98CH10886. The work at the University of Kentucky was supported by NSF through Grant DMR-1265162. Use of the Advanced Photon Source was supported by the U. S. Department of Energy, Office of Science, Office of Basic Energy Sciences, under Contract No. DE-AC02-06CH11357.

\end{document}